\documentstyle[pre,aps,multicol,epsfig]{revtex} 

\begin{document}

\draft

\author{Mario Castro,$^{1,3}$ Rodolfo Cuerno,$^{2}$ Angel 
S\'anchez,$^{2}$ and Francisco Dom\'{\i}nguez-Adame$^{3}$}

\address{$^1$Universidad Pontificia de Comillas, E-28015 Madrid, Spain\\
$^2$Grupo Interdisciplinar de Sistemas Complicados, 
Departamento de Matem\'aticas,\\  Universidad 
Carlos III de Madrid, E-28911 Legan\'es, Madrid, Spain\\ 
$^3$Grupo Interdisciplinar de Sistemas Complicados,
Departamento de F\'{\i}sica de Materiales\\ Facultad de Ciencias
F\'{\i}sicas, Universidad Complutense, E-28040 Madrid, Spain}

\title{Multiparticle Biased DLA with surface diffusion: a comprehensive model
of electrodeposition}

\maketitle

\begin{abstract}
We present a complete study of the Multiparticle Biased 
Diffusion-Limited Aggregation (MBDLA) model supplemented with 
surface difussion (SD), focusing on the relevance and effects of the
latter transport mechanism. By comparing different algorithms, 
we show that MBDLA+SD is a very good 
qualitative model for electrodeposition in practically all the range
of current intensities {\em provided} one introduces SD in the model
in the proper fashion: We have found that the correct procedure involves
simultaneous bulk diffusion and SD, introducing a time scale arising 
from the ratio of the rates of both processes. We discuss in detail 
the different morphologies obtained and compare them to the available
experimental data with very satisfactory results. 
We also characterize the aggregates thus obtained
by means of the dynamic scaling exponents of the interface height, 
allowing us to distinguish several regimes in the mentioned interface
growth. Our asymptotic scaling exponents are again in good agreement 
with recent experiments. We conclude by discussing a global picture of 
the influence and consequences of SD in electrodeposition.
\end{abstract}


\begin{multicols}{2}

\section{Introduction}
Quasi-two-dimensional (quasi-2D) electrochemical deposition (ECD) 
\cite{Meakin,Barabasi,Matsushita,Grier,Chazalviel,Marshall,Fleury%
,Leger,Leger2} has become one of the most widely studied pattern 
forming processes since
its recognition as a paradigm of {\em non-local, non-equilibrium growth
processes}\/ \cite{Meakin,Barabasi}.
Within this general context, a great deal of work has been devoted in 
the past fifteen years to experimental and
theoretical studies of quasi-2D ECD.
A first group of works deals mainly with pattern formation, its 
main results concerning ``phase diagrams'' of morphologies
\cite{Sawada,Sagues}, ECD as Laplacian growth process 
\cite{Wittenprl,Vicsekprl,Argoul,Xiao,Angel}, dynamic morphological
transitions \cite{LopezS,finger}, etc. All these studies aim to 
understanding the principles underlying the rich variety of
morphologies observed, ranging from dendritic to fractal. 
In addition to this line of research, there is a second one 
\cite{Pastor,Tesis_Juanma,Kahanda,Iwamoto,Mario,Schilardi} whose main
interest is the existence of universality and scale invariance in 
the roughness of the deposits produced \cite{Meakin,Barabasi}.
{}From all this and related research, it is now believed that
complex structures with different morphologies arise from
quasi-2D ECD due to the interplay of different transport mechanisms,
such as cation diffusion, electromigration, fluid convection, and
surface diffusion (SD)
\cite{Matsushita,Grier,Chazalviel,Marshall,Fleury%
,Leger,Leger2}.
However, the combined effect of all these factors leads to a very
complex process, and it is becoming increasingly apparent that ECD
is not well understood yet. In particular, the detailed role of
surface diffusion (SD) is still an open question that hinders our
understanding of both the morphologies and the scaling of ECD 
aggregates. 

Much of the work mentioned in the above paragraph has been 
motivated by the quest to find a universal model to help 
understand ECD phenomena. The first model formulated with that purpose was  
the famous computer algorithm known as Diffusion Limited
Aggregation (DLA) \cite{Wittenprl}, in which a particle diffuses
on a lattice and attaches to the growing aggregate at the place
where it first hits. It is not difficult to observe (see \cite{Meakin,Barabasi}
and references therein) that this simple model represents the zero 
concentration, quasi-static limit of ECD. Therefore, its validity 
as a general description of ECD is rather restricted because it 
does not include most of the effects involved in the process. However,
DLA has played a seminal r\^ole as a source of inspiration both for
continuum approaches \cite{Witten,Nauenberg,Keblinski} ---which predict some
high-current properties but take into account neither the influence
of the applied voltage nor the electrolyte concentration--- and for more
sophisticated computer models, basically modifications of DLA 
(see, e.g., \cite{Banavar,Xiao2,Erlebacher} and also the paragraph below), 
which are more or less 
phenomenological and concentrate on changes in morphology, thus being  
unable to explain the underlying mechanisms yielding
those patterns. 


In this paper, we report on the results of detailed numerical studies
of Multiparticle Biased Diffusion Limited Aggregation 
(MBDLA)\cite{Angel,Mario} supplemented with SD. MBDLA is a model in 
the family of multiparticle DLA models
\cite{Meakin,Erlebacher,Voss,Uwaha,Nagatani}, 
in which a finite number of random walkers, possibly with constant 
concentration, is introduced instead of the single walker of DLA. 
Thus, the excluded volume interaction among the walkers leads to 
several of the effects neglected in DLA. As its main ingredient,
MBDLA includes, in addition, a preferential bias (which had been first
studied in the context of single-particle,
DLA models by Meakin \cite{Meakin2})
of the walkers 
towards the cathode to mimic the electric field: In this form, 
the model 
was successfully introduced in \cite{Angel} to study the 
influence of the applied electric field on the composition of magnetic,
amorphous CoP alloys grown by ECD at constant current.
The main virtue of MBDLA is 
that it is a mesoscopic model embedded on a two-dimensional
square lattice, but it reproduces the mean fraction of Co and P in 
two-cation species ECD, as well as the qualitative morphology
of the product electrodeposits. In fact, the agreement between MBDLA and 
ECD experiments \cite{Bernal} is quantitative, as the electrical current
intensity and the experiment time can be directly related to the simulation
parameters \cite{Angel}. Therefore, we are confident that MBDLA is a good
starting point to study the relevance of SD in ECD and, specifically, its
influence on the shape of the aggregates and their dynamic scaling.
Scaling properties of MBDLA without SD were briefly reported in 
\cite{Mario}.

The report of our results is organized as follows:
We describe our model in Sec.\ II, where a brief introduction to the
physics and chemistry of ECD is followed by a detailed account of the
rules governing MBDLA. 
Section III reports on our 
numerical results, such as morphological patterns and roughness 
scaling. After physically showing that SD has to be included, we
introduce three different rules for SD are carefully considered and compared
to experiments, allowing to identify the proper way to introduce SD in
the model.  Finally, we conclude in Sec.\ IV with 
a discussion of our results which will allow us to suggest a reasonably
approximate picture 
of ECD phenomena.
A few technical details about one the rules for SD are given in an Appendix.

\section{The Model}

\subsection{Basic facts about ECD}

Prior to describing in detail what MBDLA is, and in 
order to motivate and to better understand the model rules,
we will briefly summarize the basic physics and chemistry of ECD, 
by collecting the
equations commonly accepted to govern its main features (see, e.g.,
\cite{Chazalviel,Marshall} for further details).
Generally speaking, 
ECD experiments involve two species, named cations and anions, moving in 
an incompressible viscous fluid. In very many cases, ECD takes place
in quasi-2D cells with parallel electrodes. The cations
move towards the cathode and the anions towards the anode. The basic 
equations for the concentrations of both species are as follows:
\begin{mathletters}
\begin{eqnarray}
\frac{\partial C}{\partial t}&=&-{\mathbf\nabla\cdot J}_c,\label{laplace}\\
\frac{\partial A}{\partial t}&=&-{\mathbf\nabla\cdot J}_a,\\
{\mathbf J}_c&=&-D_c{\mathbf\nabla} C+\mu_c {\mathbf E}C+{\mathbf v}
C,\label{current}\\
{\mathbf J}_a&=&-D_a{\mathbf\nabla} A-\mu_a {\mathbf E}A+{\mathbf v} A,
\label{dif-eq}
\end{eqnarray}
\end{mathletters}
where $C$ and $A$ are the cation and anion concentration respectively, 
$D_{c,a}$ the cationic and anionic diffusion coefficients, $\mu_{c,a}$ their
mobilities, $\mathbf{v}$ the fluid velocity field, and $\mathbf{E}$ 
the electric field along the cell. The latter 
is related to cation and anion
concentration via the Poisson equation
\begin{equation}
{\mathbf\nabla\cdot E}=-{\mathbf\nabla}^2\phi  
=-e(z_cC-z_aA)/\varepsilon,
\end{equation}
where $\phi$ is the applied potencial, $ez_c$ and $-ez_a$ are the cation
and anion 
electric charges, respectively, and
$\varepsilon$ the dielectric permittivity of the fluid.
Generally speaking, matter balance across the interface leads to an 
interface velocity proportional to 
the flux of cations 
and therefore, in the absence of any other limiting process, 
proportional to the
current density as well. In addition,                             
except for the
region close to the cathode, 
we may assume electro-neutrality \cite{Chazalviel}, which in turn implies
that the cation mean velocity is constant.
The incompressible Navier-Stokes equations determine
the velocity $\mathbf{v}$ of the solvent. 
Fluid convection is always present in ECD experiments, but
in many instances it
can be small enough to be safely 
neglected, as has been shown in actual experiments \cite{Huth,Chazalviel2}.

When the cations arrive at the cathode, they  reduce irreversibly and
an aggregate of neutral particles begins to grow. The particles
on the surface aggregate are transported all along it due to
local chemical potential
gradients. The resulting particle current conserves the number of
particles on the surface and is
given approximately by by (see \cite{Mullins} for a detailed
discussion): 
\begin{equation}
{\mathbf J}_{s}\propto -{\mathbf\nabla}_s\kappa,
\end{equation}
where ${\mathbf J}_{s}$ is the particle current along the surface, 
$\kappa$  the 
interface local curvature at each site, and
${\mathbf\nabla}_s$ the gradient taken
along the surface. 
Roughly speaking, SD tends to reduce the interface local curvature.
Finally, we note that the
mean concentration of charge carriers in the bath is constant as
new cations are formed at the anode upon arrival of the anions
\cite{Fleury}.

\subsection{Definition and rules of MBDLA}

In this section we will define MBDLA through its
evolution rules, for which we take into account the
physical equations presented in the previous section.
At this point, we do not consider SD, whose need will be justified
in the next Section, and consequently we postpone the discussion of
the rules to implement SD as well.
Thus, MBDLA is a {\em cellular automaton} 
defined on a two-dimensional square lattice of 
horizontal dimension $L_x$ and vertical dimension $L_y$ (with lateral 
periodic boundary conditions and reflective boundary condition
at the top; for the conditions at the bottom, see below),
in which a number of random walkers ({\em cations})
are randomly distributed with concentration $c$. The bottom of the lattice
is chosen to be the cathode. We do not consider the anion dynamics, but we  
implicitly introduce it by the creation of particles and by charge
electro-neutrality \cite{foot}.

The initial condition evolves in time
as follows: Every time step a walker is chosen
and moved to one of its four neighboring sites with 
probabilities taken from a finite differences scheme of 
Eqs.\ (\ref{laplace}) and (\ref{current}) \cite{Huang}: 
probability $1/(4+p)$ to move either left, right
or upwards, and probability $(1+p)/(4+p)$ to move down, i.e.,
towards the cathode.
The parameter $p\/$ is referred to as the {\em bias}; in galvanostatic
conditions it can be
quantitatively related to the electric current density in the physical
system as shown in \cite{Angel}. Let us stress here that our present 
choice for the probabilities is different from that reported in 
\cite{Angel} and \cite{Mario}, but we have checked that
the results hardly differ with those presented in this paper. 
The main reason for this new election is that, with the new rules,
the bias $p$ ranges from $0$ to $\infty$, that is, from pure multiparticle
DLA to ballistic
deposition, whereas the rules in the mentioned references allow for a
range in $p$ from $0$ to $0.25$, and the ballistic deposition limit 
cannot be reached (although $p=0.25$ is rather close already, see 
\cite{Angel,Mario}). 
After a destination site has been chosen, the particle
moves if that node of the lattice is empty; if not, we select another
particle and repeat the destination selection procedure. 
Once the particle has been moved, if the new position has
any nearest neighbor site belonging to the aggregate, the
present position of the walker is added to the aggregate (the cathode
or bottom boundary at the initial stage) with
probability $s\/$ (and is able to diffuse over the aggregate surface if
that aggregate position has just one nearest neighbor belonging to it, see
the following section); otherwise it stays there (and is able to move again)
with probability $1-s$.  We term $s\/$ the {\em sticking probability}; it
is related to the chemical activation energy the cation needs to stick to
the aggregate.
As particles are added to the aggregate, 
other ones are created at the top of the lattice keeping the mean cation
concentration $c$ constant, which in fact simulates an
infinitely high system (experimentally this means that the distance between 
electrodes is much larger than their lateral dimension); consequently, 
the flux of particles is constant at every stage of the simulation.

As we have already pointed out, the model parameters are related to 
the physical factors influencing the problem. Indeed,
the choice of jump probabilities for the random walkers in the bath
provides a recurrence relation which is a discretized version of the
continuous
equations (\ref{laplace}) and (\ref{current}). 
Therefore, the drift velocity $\mathbf{\mu_c{\mathbf E}}$ is
proportional to 
the {\em bias} $p$. When
a finite number of walkers is considered with concentration $c$, we
must take into 
account the excluded volume, so the effective diffusion coefficient and
the 
effective drift velocity in the simulations are proportional to $1-c$
(in a mean field approach) \cite{Comentario}.
It is important to note that  when $c\rightarrow 0$, {\em i.e.}, the bath
is formed by one particle alone (as in DLA), the aggregate develops
tall branches which grow at the expense of short ones due to screening 
effects. Therefore, in the low current limit a morphological instability 
appears that is not always present in ECD experiments.
The finite concentration and the
hard core interaction among random walkers simulates the
cation {\em pressure} on the
aggregate, so $c$ is an essential ingredient in the understanding of the
formation of electrodeposits and to prevent these instabilities (of 
Laplacian character) from dominating the whole growth process. 

One important task is the definition of simulation time step.
In \cite{Angel}, comparison with the experiments in \cite{Bernal}
allowed to show that the physical time and the 
simulation time measured in number of Monte Carlos trials were
simply proportional to each other. 
For this reason, we have stuck to the definition of 
the time step in \cite{Angel} as a Monte Carlo trial, {\em i.e.},  
the time needed for a particle to jump, both if the particle
does jump or if it does not. Notwithstanding,
we have tried other time steps definitions, such as the Monte Carlo step
being defined as the mean time for every random walker to jump
at least once, but the results are basically the same. 
Some authors define the time step for solid-on-solid
growth models as the mean time
needed to complete an aggregate layer, but
as we will show below, ECD electrodeposits do not grow with constant
velocity, and therefore 
the mean interface height does not grow linearly with time. We thus 
believe that, in the ECD context, this time unit would be rather 
artificial and hence we have not used it.  In fact, as we will show
below, the work reported in this paper provides further evidence in
favor of our choice (see the discussion of the experiments in
\cite{Schilardi} in Sec.\ III B below).

\section{Numerical Results}

\subsection{Morphologies }

We begin the summary of our results by discussing the morphologies generated
by MBDLA with and without
SD and the influence of the different rules for SD on them. In addition, 
we want to compare our computer generated morphologies to the available 
experimental data. We take as a reference the comprehensive experimental 
work of Trigueros {\em et al.} \cite{Sagues}, who reported a systematic 
experimental study of different growth regimes at constant applied voltage
conditions. Their work gave rise to 
a diagram of morphologies divided into different regions in
which similar morphologies were obtained
as a function of the applied voltage and the electrolyte concentration.   
It is important to realize that, in galvanostatic conditions, there is no
linear correspondence between voltage and electric current of ions, and
therefore, comparison between our morphologies and those reported by
these authors can only be qualitative. No similar taxonomy work has been
performed for constant current conditions. 
Although the diagram in \cite{Sagues} is quite complex, it 
encloses a full variety of morphologies under the label {\em compact}.
Some authors \cite{Pastor,Kahanda} have studied electrodeposits systems
within this regime, and hereafter we will
also refer to them. Finally, a recent work by Schilardi {\em et al.}
\cite{Schilardi} provides exhaustive information on the asymptotic
ECD regimes, which have not been considered anywhere else; hence, 
their research will also be compared to ours along the paper.

As we have already mentioned, from the model perspective we can compare the 
bias $p$ with the electric current density \cite{Angel,Comentario_p}, 
and $c$ with  the electrolyte concentration, even though the two latter 
magnitudes are not exactly coincident, i.e., an electrolyte concentration 
equal to $0.1$  M does {\em not} mean $c=0.1$. We will see below that the 
results are not very sensitive to the specific value of $c$ insofar it is 
not very small, and thus the difference between actual and model 
concentrations is not very relevant. The sticking probability, $s$, and 
the diffusion parameters, namely $l$, $\lambda$ and $r$ (or equivalently 
$\tau_d$), cannot be directly tuned in an experiment, 
although it is reasonable to 
expect that changes in the experimental conditions will correspondingly modify 
these parameters. How much are they modified is something we will learn 
through our computer simulations.   

\subsubsection{Bias vs. Sticking Probability without SD}

\begin{figure}
\begin{center}
\epsfig{file=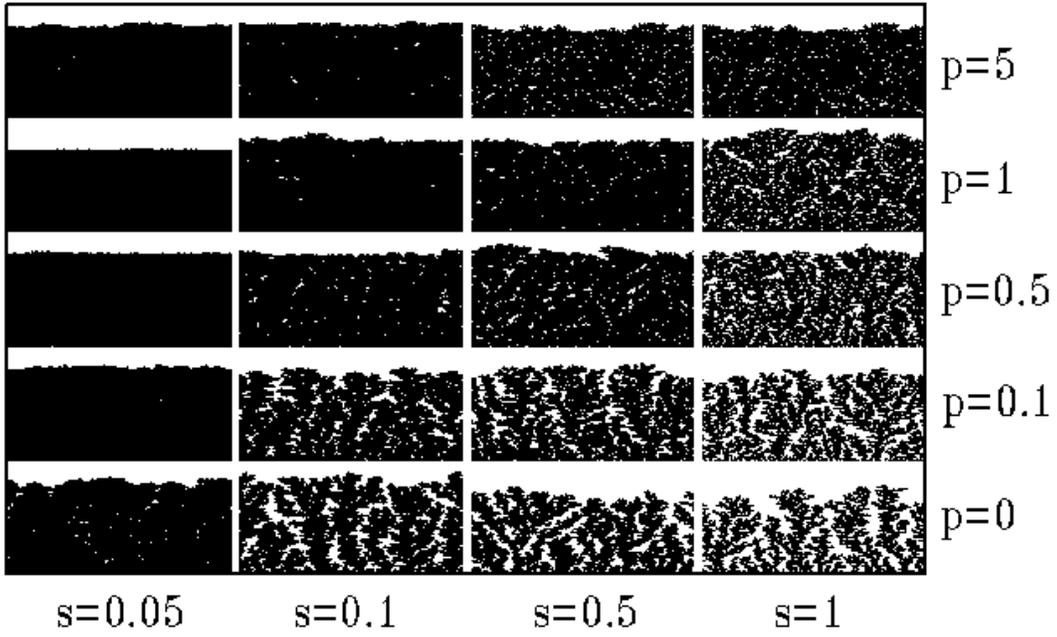, width=5.6in, angle=0}    
\end{center}
\caption[]{Morphologies obtained with MBDLA without surface
diffusion for $256\times400$ systems with a
cation concentration $c=0.05$. Other parameters are as indicated
in the figure.}
\label{pdes}
\end{figure}

Figure \ref{pdes} shows a diagram of morphologies obtained with 
$0\leq p\leq 5$ and $0.01\leq s\leq 0.5$ {\em without SD}, with a particle 
concentration $c=0.05$. We have included these results for two reasons:
First, there has been no previous report on MBDLA morphologies, except for
a brief discussion in \cite{Angel}; and, second, we need to discuss them
in order to understand later what is the effect of SD on MBDLA morphologies. 
It is clear from Fig.\ \ref{pdes} that 
increasing the bias or decreasing the sticking probability yields
denser aggregates, the ones obtained for $p=0$ and $s=1$ (bottom 
right) being multiparticle DLA-like as expected (compare to \cite{Voss,Uwaha}).
This phenomenon is related to the 
stabilizing effect of the parameters $p$ and $s$, which can be theoretically
demonstrated \cite{Futuro}. Indeed, the higher the values of $p$, the
larger the flux of particles reaching the interface in the direction 
perpendicular to the cathode. This reduces the probability for a cation 
to stick laterally to a branch and the screening effects due to the 
Laplacian field. On the other hand, the electric field combined 
with the reduction of the sticking probability 
tends to fill the interface valleys.
This first result, namely the fact that increasing the electric current 
leads to denser aggregates, is similar to the results reported 
by Trigueros {\em et al.}\ \cite{Sagues}, who observed densification
of the aggregates with increasing applied voltage.
In particular, we can qualitatively compare the morphological changes
obtained by varying the bias $p$ for a fixed $s=0.5$ in Fig.\ \ref{pdes}, 
with those provided by experimental voltage variations 
(see Fig.\ 2 in \cite{Sagues}). We conclude that
high voltages (or in general, high density currents) yield denser
aggregates.
So the bias $p$ is an essential ingredient in any realistic ECD model.

\begin{figure}
\begin{center}
\epsfig{file=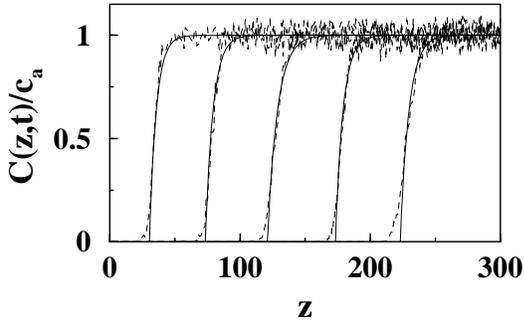, width=2.7in, angle=0}    
\end{center}
\caption{Concentration profiles for a $512\times 300$ system with
parameters $p=1$, $s=1$ and 
$c=0.1$ without surface diffusion. Dashed lines represent the
simulation data and solid
lines the best fit of those data to Eq.\ (\ref{density_fit}).
The height $z$ is given is
lattice spacings.}
\label{concentration_density}
\end{figure}

\begin{figure}
\begin{center}
\epsfig{file=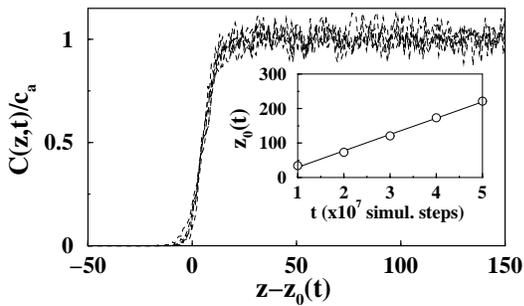, width=2.7in, angle=0}    
\end{center}
\caption{Collapsed concentration profiles using the values of $z_0$
obtained from
Eq.\ (\ref{density_fit}). Inset: The mean concentration front
position, $z_0$, {\em vs} time.}
\label{collapse_density_fit}
\end{figure}

As a second step in our study, 
we have monitored other relevant quantities 
which in turn can be experimentally measured, in order to obtain additional
information aside from qualitative morphological comparisons.
Figure \ref{concentration_density} 
shows the local concentration of particles
in the bath, still without  SD, at equal time intervals.
We have plotted the concentration profiles in the stationary
regime, {\em i.e.}, after the
instability occurs (see below). 
Thus, the mean number of attached particles per unit time (or
equivalently, the mean interface velocity) is constant. L\'eger {\em et al.}
\cite{Leger,Leger2} have reported experimental
evidence consistent with this stationary behavior
(see {\em e.g.} Fig.\ 5 in \cite{Leger2}). We thus see 
that  MBDLA agrees well with their 
findings, i.e., 
the stationary concentration of particles 
in the bulk obeys approximately the equation 
\cite{Leger2,Futuro}:
\begin{equation}
C(z)=c_a+(c_0-c_a)e^{-(z-z_0)u/D_c},
\label{density_fit}
\end{equation}
where $z$ is the vertical coordinate, $z_0$ is the  interface mean position, 
$c_a$ is the concentration at the anode,
$c_0$ the concentration at the interface, $u=\mu_c{\mathbf E}$,  
and $D_c$ is the bulk diffusion coefficient. As shown in Fig.\
\ref{concentration_density},
this function provides a good fit of our data. 
In Fig.\ \ref{collapse_density_fit}, we
plot a fit of Eq.\ (\ref{density_fit})
(dashed line) to the simulation results, 
showing a good collapse of the bulk particle density
outlines for different times. 
The small deviations close to the aggregate are due to the 
interface roughness.
The ratio $D/u$ is called diffusion length. In our fits, this length 
turns out to be
about $15$ lattice spacings, that is, about $2$ or $3$ times the lateral width
of the branches for the chosen parameters. This result provides another
of check the physical validity
of our model, as we can compare the length obtained from the fit
with that taken from Ref.\ \cite{Leger2}.
In this paper, the diffussion length is  of order $0.5$ $mm$, about two
times the typical branch lateral width (of order $1$ $mm$), so we may
conclude that the diffusion length obtained from our model 
is physically consistent.

The inset in Fig.\ \ref{collapse_density_fit} shows
the mean concentration front
position $z_0$ {\em vs.} time, demonstrating that, in the 
stationary state, MBDLA leads to 
a constant velocity of the advancing front as in the experiments.

\subsubsection{Physical Relevance of SD}

The previous subsection shows that MBDLA without SD successfully
reproduces some 
ECD experiments, in particular, under galvanostatic conditions
with not very small electric 
current density. However, within the MBDLA model it is impossible 
to understand 
the unexpected compactification of aggregates in 
low voltage experiments \cite{Sagues,Pastor,Kahanda}
or the columnar-like growth found in other situations.
Unfortunately, MBDLA aggregates are always ramified at low bias. In 
\cite{Angel}, a phenomenological explanation of compactification was
proposed by noticing that the reduction of $s$ leads to more compact
aggregates. Therefore, it was proposed there that $p$ and $s$ should 
be related by a monotonous function, the simplest case being that of
a linear relationship. With this procedure, reducing the
bias leads to a corresponding decrease in the sticking probability, 
and hence to compact aggregates at low bias. 
However, this is an {\em ad hoc} assumption that 
cannot be experimentally tested, whereas its theoretical
justification is not very clear. 
Besides that, this approximation does not reproduce other morphologies, 
as that reported by  L\'opez-Salvans {\em et al.}
\cite{finger} or Kahanda {\em et al.} \cite{Kahanda}.
In view of this, it became increasingly clear that there was some 
crucial ingredient missing in MBDLA, and the most obvious candidate 
was of course SD. 

At this point, it is instructive to consider carefully 
the work by Kahanda {\em et al.} \cite{Kahanda}. 
According to their results, as the 
absolute value of the overpotential
decreases the aggregate becomes denser, and it is formed by
several columns which are thicker at the top than at the bottom. 
We interpret this as a hint on the relevance of 
SD: If, when a particle arrives at the top
of a column, it diffuses along the aggregate interface, and if the diffusion
length is shorter than the column perimeter, the particle will not reach the
base of the pillar or another column, with the result of
a characteristic inverted triangle structure. The onset of similar triangle
structures has also been reported by Pastor and Rubio \cite{Pastor}.
We thus came to the conclusion that it was necessary to include 
SD in MBDLA in order to shed further light on the nontrivial coupling of
the different transport mechanisms. 

\subsubsection{Implementation of SD in MBDLA}

We have implemented SD in MBDLA in three different ways, all of them 
starting when a particle in the {\em bulk} (the electrolytic solution)  
sticks to the aggregate but has just one neighbor. We have first
tried two simple irreversible rules (other similar rules yield
equivalent results, so we do not include them here for brevity), named 
rules A and B, and an irreversible one, named rule C:

{\bf Rule A:}
The newly incorporated particle jumps always in the same direction, 
either left or right parallel to the cathode, until it reaches a 
site with at least $2$ neighbors or completes $l$ jumps. This rule is similar,
but not identical, to the one studied in \cite{Yan} for ballistic 
deposition with surface diffusion.

{\bf Rule B:} In this second rule, we allow the particle to perform a
random walk over the aggregate surface until it increases its
coordination number, with a constant probability $\lambda$ to be
permanently stuck to its current position (this is the 
so called {\em mortal random walker}\/ \cite{Hughes}).

The last rule is characterized by
Arrhenius-like jump probabilities and, what is more important,
by {\em simultaneous bulk diffusion and SD}:

{\bf Rule C:}
This rule allows several particles to diffuse simultaneously. When a particle 
arrives to a coordination $1$ site, it sticks and jumps to one of its two 
nearest neighbors on the aggregate with probability 
$p_n=\exp[-E_0+(n-1)E_a]$, where $E_{0,a}$ are adimensional activation 
energies, and $n$ the coordination number of the target position.   
If the particle new position has $2$ or $3$ neighbors, it attaches to 
the aggregate irreversibly. Otherwise, we {\em label} the particle as a 
SD particle, and we allow it to take further steps. Thus, we have two kinds 
of diffusing particles: Particles in the bulk, distributed homogeneously with 
concentration $c$; and particles that diffuse over the aggregate surface. 
With probability $r$ we choose a bulk particle which evolves with its 
characteristic rules, and with probability $1-r$ a particle on the surface 
which jumps to one of its nearest neighbors as we have just described 
for the first jump. This rule is close in spirit to the collective diffusion
rules employed in studies of kinetic roughening in molecular beam epitaxy (MBE)
\cite{Krug}, and in particular to MBE models beyond the solid-on-solid
approximation \cite{Das Sarma,Schimschak}.

The main difference between rules A and B with respect to rule C
is that the latter introduces a characteristic time scale $\tau_d=r^{-1}$,
while in the other cases diffusion is 
instantaneous; then, the diffusing particle is not affected by the
overall particle dynamics. 
As we will show below, Rule C is the only one which actually reproduces
the influence of SD on the aggregates scaling and 
morphology. In this respect, it is important to advance that we 
have found that Arrhenius-like 
probabilities by itself are not enough to model SD: Variants of rule
C with those probabilities and without the characteristic time, i.e.,
SD kept instantaneous, lead once again to results similar to those of rule B.
All the results presented were obtained with $E_0=3$ and $E_a=1$.
We have chosen these values to have jump probabilities smaller
than $1$, but other sets of parameters yield similar results
which we omit for brevity. Finally, another interesting point is
that the probabilities in Rule C allow to trivially introduce 
temperature in the model by simply identifying
$E_{0,a}\rightarrow E^\prime_{0,a}/k_BT$.

\begin{figure}
\begin{center}
\epsfig{file=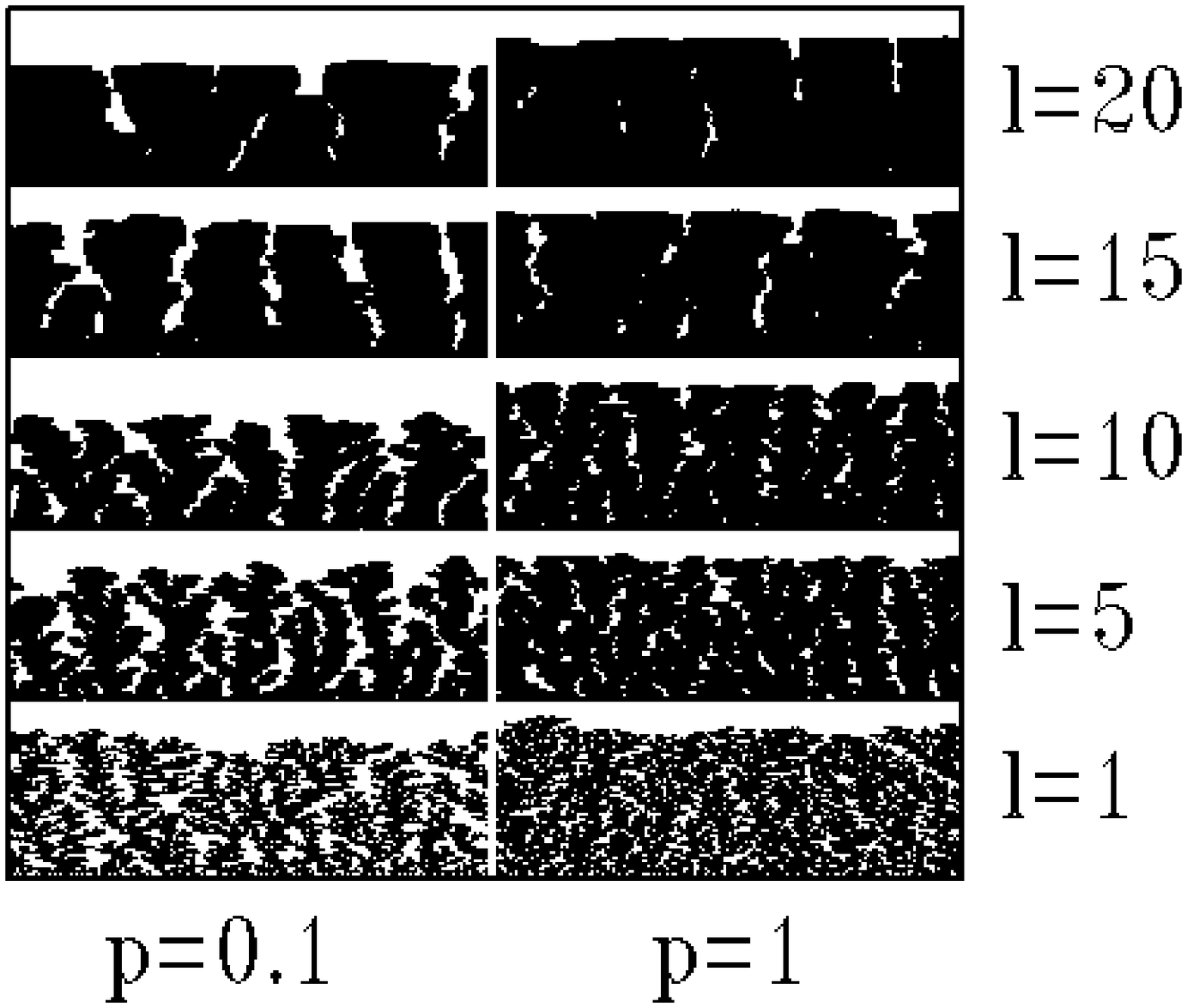, width=2.7in, angle=0}    
\end{center}
\caption{Morphologies obtained with MBDLA with surface diffusion rule A
and parameters $s=1$ and $c=0.1$. The size of the system is
$256\times400$ pixels.}
\label{rule_A}
\end{figure}

\begin{figure}
\begin{center}
\epsfig{file=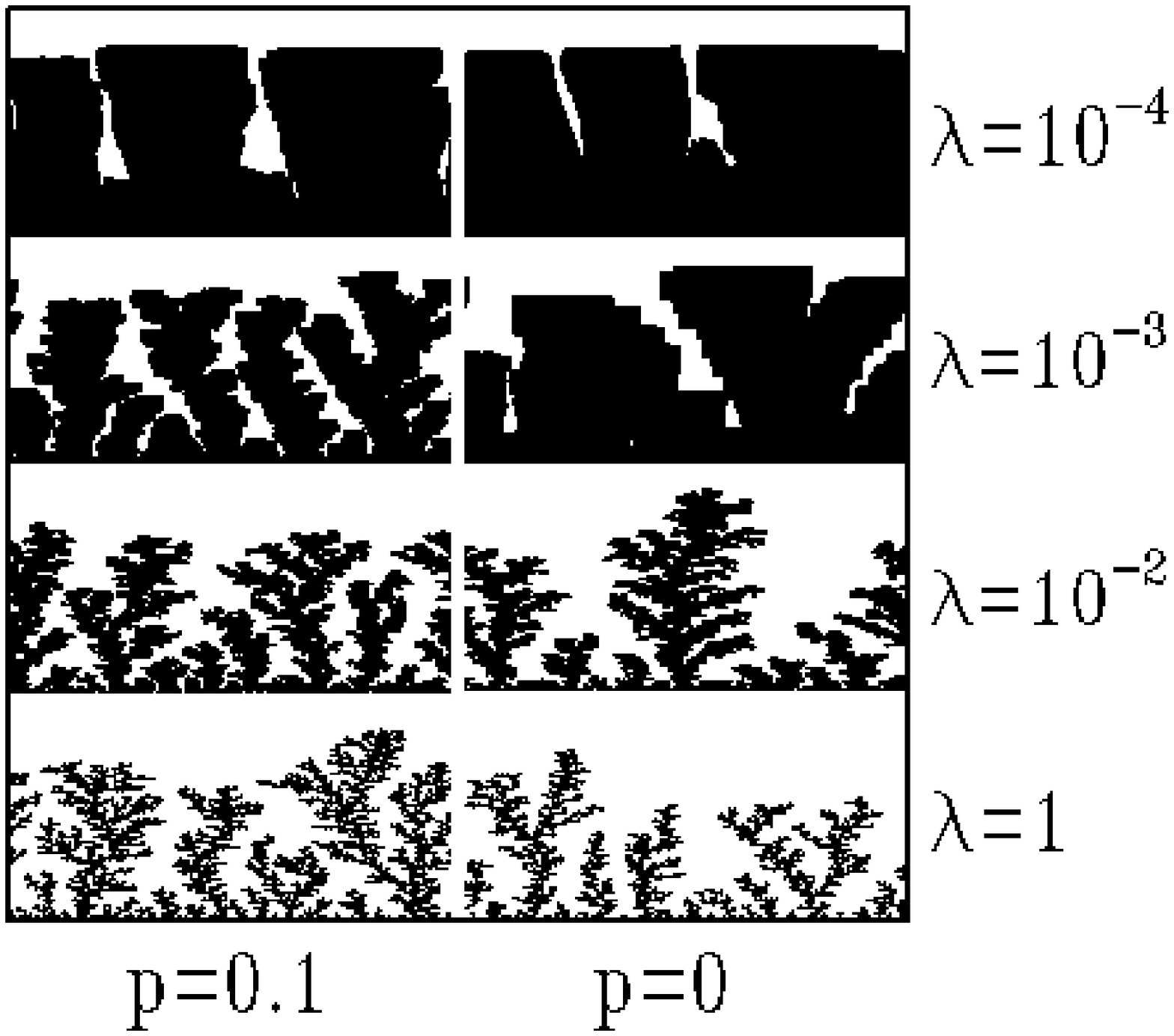, width=2.7in, angle=0}    
\end{center}
\caption{Morphologies obtained with MBDLA with surface diffusion rule
B 
and parameters $s=1$ 
and $c=0.1$. The size of the system is $256\times400$ pixels.}
\label{rule_B}
\end{figure}

Rule A, by definition, introduces a diffusion length $l$, but
if $l\gg 1$ the particle jumps practically always lead to an increment of
its coordination, as may be seen  in Fig.\ \ref{rule_A}, where
some morphologies are shown for different values of $l$. 
The inverted triangle structure typical of the experiments by 
Kahanda {\em et al.}\ \cite{Kahanda} is reproduced
with this simple rule. Nevertheless, the tops of the pillars are
unrealistically flat; another problem is that decreasing $p$ 
does not lead yet to a compact aggregate regime. Rule A is therefore
not appropriate. 
In the case of Rule B, 
the diffusion length is introduced indirectly  by means of the 
attachment probability 
$\lambda$ (see Appendix for details).
The mean diffusion length can be shown to be given by
$l_D=1/(2\lambda^{1/2})$. 
Morphologies obtained with this rule are plotted in Fig.\
\ref{rule_B}. Once again, and
in spite of the fact that rule B allows the particles to
diffuse randomly over the
aggregate, 
the columns developed during the growth turn out completely
flat at the top, and the option of rule B was excluded as well. 

\begin{figure}
\begin{center}
\epsfig{file=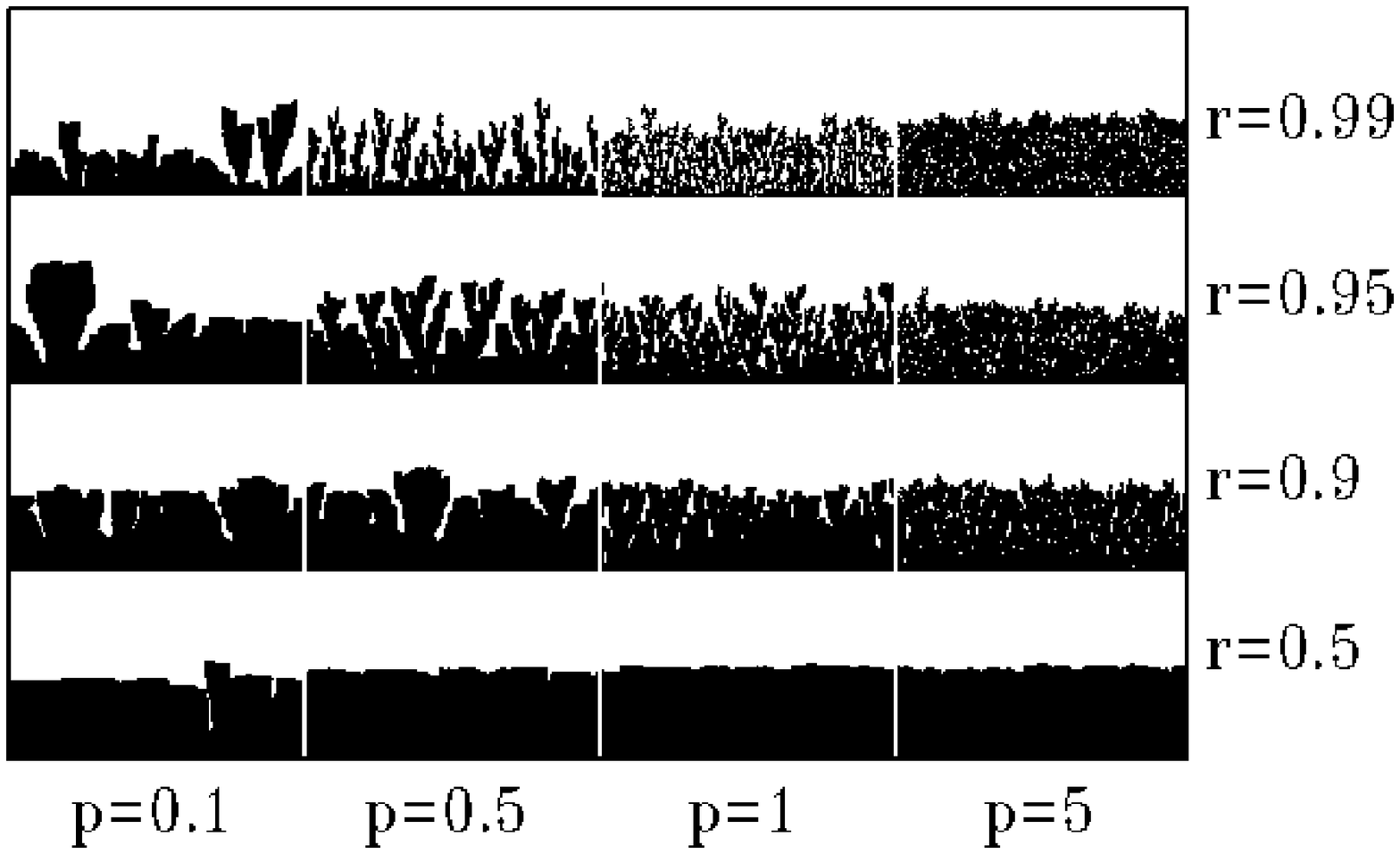, width=2.7in, angle=0}    
\end{center}
\caption{Morphologies obtained with MBDLA with surface diffusion
rule C and
parameters  $s=1$ and $c=0.1$. The size of the system is
$256\times400$ pixels.}
\label{rule_C}
\end{figure}

These pitfalls (and similar ones found by using Arrhenius-like jump
probabilities, which we skip for brevity) led us to the conclusion that 
instantaneous SD (or limited mobility rules, in the terminology of \cite{Krug})
is a too drastic one approximation for ECD. Taking into account 
that the main distinctive feature of MBDLA is its non-local character, 
interactions of particles diffusing along the surface with newly deposited
particles are expected to be relevant. Guided by these ideas, we propose 
rule C, which incorporates this coupling by introducing time scales for
both bulk diffusion and SD. A sample of the aggregates generated by 
MBDLA with rule C is shown 
in Fig.\  \ref{rule_C}. The difference with the other two rules is 
immediately apparent from the plot: This
more realistic rule does induce the creation of pillars as we pointed 
out above, this time
similar to those reported by Kahanda {\em et al.} \cite{Kahanda} and 
Pastor and Rubio \cite {Pastor} which are rough at the top.  
Moreover, the compactification of the aggregates at low currents 
appears naturally, as can be noticed by following the sequence of 
aggregates appearing on the same row (same value of $r$): Decreasing 
the current leads initially to less dense aggregates, until
further reduction of the current gives rise to more compact aggregates.
Remarkably, there is no need to change the sticking probability 
by hand as in MBDLA without SD or with Rules A and B.
This allows us to eliminate one model parameter, the sticking probability,
which we take to be $s=1$ from now on.

So far, we have seen 
that, while simple SD rules provide
good results in some solid-on-solid simulation models, the complex 
dynamics of Laplacian systems does not allow the particles to 
instantaneously diffuse; rather, we must allow several particles to
interact before they become permanently stuck to the aggregate.
Roughly speaking, the flux of particles arriving at the aggregate
defines a characteristic time 
$\tau_p$ (typically inversely proportional to the flux, {\em i.e.}, to $p$). 
Once the particles have arrived to the aggregate, they diffuse until they 
reach a site with coordination larger than one, or equivalently, until the
particle meets another diffusing particle, thus forming a dimer
on the interface which cannot move anymore. 
A large flux of particles arriving at the interface (large $p$) will 
increase the probability of formation of those dimers, and the particles 
can hardly diffuse.   
The situation is not so simple when $p$ is small. On one
hand, the deposition mean time $\tau_p$ is large, but on
the other hand, the particles hardly experience 
the applied electric current, so the probability of attachment to a
column wall before getting to the bottom of the aggregate increases.
Thus the Laplacian instability is
amplified leading to a compact structure 
formed by columns and grooves. This kind
of instability has been observed in low current galvanostatic experiments
\cite{Pastor}.
The aggregate is therefore denser but if the diffusion time
is not long enough the interface is unstable. 
It is remarkable that this simple picture in terms of
time scales allows to understand the relevance of SD in 
ECD experiments.

A final important remark we would like to mention 
is that, when the diffusion probability 
$r$ is about $0.99$, we have observed some evidences 
of what could be a morphological
transition (and the subsequent change in the branches) similar
to those reported by L\'opez-Salvans {\em et al.} \cite{LopezS}.
However, as we want to concentrate in this paper on MBDLA
with SD as a generic model for all regimes of ECD experiments, 
we postpone a more careful study of this possibility to future 
work, where we will pursue the appearance of this phenomenon 
for different 
model parameters (such as $p$ or $r$).

\subsubsection{Electrolyte concentration}

\begin{figure}
\begin{center}
\epsfig{file=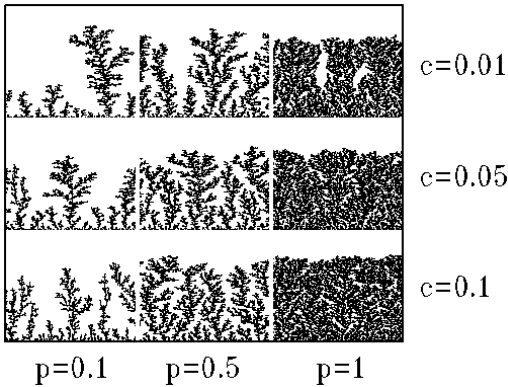, width=2.7in, angle=0}    
\end{center}
\caption{Morphologies obtained with MBDLA without surface diffusion
for parameters
$s=1$ and $r=1$. Other parameters are as indicated.}
\label{c_morph_p}
\end{figure}

To conclude the analysis of MBDLA parameters, we show  the effect of the
electrolyte concentration, $c$. Figure \ref{c_morph_p} exhibits the
morphological changes in patterns with different $c$ values ranging
from $0.01$ to $0.1$ for different bias $p$ without SD. 
Note that when $c\rightarrow 0$ the low current limit is exactly the DLA 
growth model \cite{Witten}. Therefore, we should keep a finite value of $c$ in
order to diminish the unavoidable DLA characteristic instability.  The 
results contained in the figure allow us to conclude that, insofar $c$ is 
not very small, the morphologies obtained with MBDLA do not depend 
strongly on the concentration, and therefore the fact that there is 
no direct correspondence between physical and simulated concentrations
is not a drawback of the model. 

\subsection{Dynamic Scaling}

The previous subsection shows that 
the inspection of the morphologies is a valuable method to check the
validity and relevance of the model rules. Indeed, the
unrealistically flat aggregates obtained with diffusion rules A and
B unable them and motivate the investigation of the more realistic,
non-instantaneous rule C for SD.  However, in order 
to exploit the main virtues of MBDLA with SD and to compare with other
relevant models and experiments, we must take some quantitative
criteria, for example,
the analysis of the interface surface roughening.
To this end, 
let us define some functions related to  
the height of the aggregate at spatial position $x$ at time $t$, given by
the scalar field $h(x,t)$. We will also review their basic features 
before discussing MBDLA properties.

The global width (or {\em roughness}) $W(L,t)$ is nothing but 
the rms fluctuations of the height variable $h(x,t)$ around its mean value
$\bar{h}_L(t)= (1/L) \sum_x h(x,t) $:
\begin{equation}
W^2(L,t) = \frac{1}{L} \bigg \langle \sum_x \left[h(x,t) - \bar{h}_L(t)\right]^2 \bigg \rangle,
\label{Wtot}
\end{equation}
where angular brackets stand for ensemble average.

Generally speaking, in many growth models,
starting from $h(x,0)=0$ the width satisfies the dynamic scaling
hypothesis of Family-Vicsek \cite{Family}:
\begin{equation}
W(l,t) \sim \left\{ \begin{array}{lll}
t^{\beta} & {\rm if} & t \ll L^z, \\
L^{\alpha} & {\rm if} & t \gg L^z. \\
\end{array} \right. \label{Wfv}
\end{equation}
The roughness exponent $\alpha$, the dynamic exponent $z$, and their
ratio (growth exponent)  $\beta=\alpha/z$, identify the universality class 
the model belongs to. 

In the study of kinetic roughening the height-height correlation function
is frequently used \cite{Barabasi}: 
\begin{equation}
C^2(l,t)=\frac{1}{L}\bigg\langle\sum_x \Big(h(x+l,t)-h(x,t)\Big)^2\bigg\rangle,
\end{equation}
where \cite{Schroeder,Lack},
\begin{equation}
C(l,t) \sim \left\{ \begin{array}{lll}
t^{\beta} & {\rm if} & t \ll l^z, \\
t^{(\alpha-\alpha_{loc})/z} \, l^{\alpha_{loc}} & {\rm if} & t \gg l^z, \\
\end{array} \right. \label{wlfv}
\end{equation}
where $\alpha_{loc}$ is the so called local roughness exponent.
Another important 
function related to the height variable $h$ is the power spectrum:
\begin{equation}
S(k,t) = \langle \widehat{h}(k,t) \widehat{h}(-k,t) \rangle,
\end{equation}
 where $\widehat{h}(k,t) = L^{-1/2} \sum_x [ h(x,t) - \bar{h}_L(t)] \exp({\rm
i}kx)$. 
$S(k,t)$ displays a behavior consistent with the scaling form \cite{Versus}
\begin{mathletters}
\label{Sanom}
\begin{equation}
S(k,t)=k^{-(2\alpha+1)}s(kt^{1/z}) , 
\label{Sanom1}
\end{equation}
where
\begin{equation}
s(u)=\left\{\begin{array}{lll}
u^{2\theta} & {\rm if } & u\gg 1,\\
u^{2\alpha+1} & {\rm if } & u\ll 1.\\
\end{array} \right.
\label{Sanom2}
\end{equation}
\end{mathletters}
The exponent $\theta$ takes different values depending on the type of 
scaling exhibited by the model. For instance, for the so called 
intrinsic anomalous scaling \cite{Versus} we have
$\theta=\alpha-\alpha_{loc}$, whereas  $\theta \equiv 0$ for 
Family-Vicsek scaling (including super-roughening, i.e., $\alpha\geq 1$). 
Note that this implies $\alpha=\alpha_{loc}$.

To apply these ideas to MBDLA characterization, a few remarks are in
order. 
Although, in some cases, 
MBDLA develops ramified aggregates leading to  multivalued interfaces,
{\em i.e.}, interfaces with overhangs, it has been demonstrated \cite{Vicsek}
that the interface of the active zone in DLA simulations (the aggregate sites
with larger probability of arrival) corresponds to that constructed 
by taking the topmost site $h(x,t)$ at every horizontal position $x$. 
This construction {\em does not} ensure that the measured
exponents are free of interpretations \cite{Comment}, but the
exponents are consistent with theoretical and experimental data 
\cite{KahandaReply}. The reduction of the sticking probability $s$
yields denser aggregates, and overhangs do not appear at any stage of the
simulation for low $s$ values. 
Besides that, if SD is present the aggregates are also more compact.
In all these cases the function $h(x,t)$ is identical to the aggregate outline 
and consequently the results do not have any interpretation problem.

The main scaling features of  
MBDLA without SD were already reported in \cite{Mario}. Therefore, 
here we will briefly summarize them to facilitate comparison with 
results including SD, and refer the reader to \cite{Mario} for 
the details. Without SD,
MBDLA displays three temporal regimes: 
At early times the global width $W(L,t)$ features $\beta=0.5$,
this value being simply due to shot noise. This stage corresponds to 
times at which the lateral correlation length is of the order of
the lattice spacing. After this noisy transient, short
and large length scales are governed by different 
dynamics because the bulk
Laplacian field produces nonlocal effects (screening or shadowing
among branches). 
Consequently, the local and the global roughness exponents,
$\alpha_{loc}$ and $\alpha$,
are different and the interface is not self-affine.
The growth exponent, $\beta$, is larger than that of noise ($\beta>1/2$)
because some isolated branches begin to grow independently from each other,
which can be understood as a signature of the Laplacian instability.
As a consequence, the interface width grows rapidly as compared with the noise 
fluctuations. 
At later times, 
branches spread by lateral growth and impinge upon each other. Eventually,
the system reaches an asymptotic regime
characterized by the Kardar-Parisi-Zhang (KPZ)
universality class \cite{Kpz} exponents 
($\alpha=1/2$, $\beta=1/3$, $z=3/2$). The KPZ equation is 
the paradigmatic growth model without SD, and it is given by
the stochastic partial differential equation \cite{Kpz} 
\begin{equation}
\frac{\partial h}{\partial t}=\nu\nabla^2h+
\frac{\lambda_0}{2}\left(\nabla h\right)^2+
\eta(x,t),
\end{equation}
where $\nu$ and $\lambda_0$ are constants
and $\eta(x,t)$ is a Gaussian white noise
with:
\begin{mathletters}
\begin{equation}
\langle\eta(x,t)\rangle=0,
\end{equation}
\begin{equation}
\langle\eta(x,t)\eta(x^\prime,t^\prime)
\rangle=2D\delta(x-x^\prime)\delta(t-t^\prime).
\end{equation}
\end{mathletters}

\begin{figure}
\begin{center}
\epsfig{file=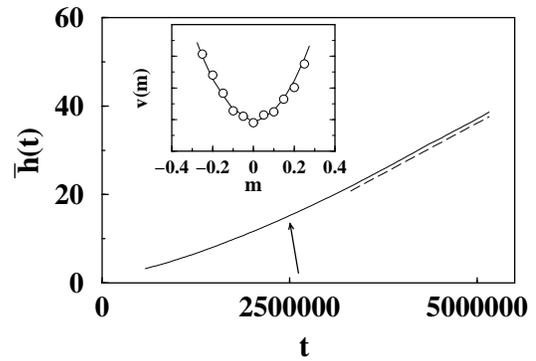, width=2.7in, angle=0}    
\end{center}
\caption{Interface mean height {\em vs} time with
parameters $p=4$, $s=1$ and $c=0.05$ without 
surface diffusion. The arrow shows the end of the unstable regime.  
Inset: Mean excess velocity, in arbitrary units, for the same parameters for different 
boundary tilts, $m$. Circles stand for simulation and the solid line is
the best fit to a parabola. The dashed line represents the expected 
linear growth of KPZ type.}
\label{meanh}
\end{figure}

As mentioned above, 
the definition of the interface function $h(x,t)$ neglecting 
overhangs might cast some doubts \cite{Comment} 
on the validity of the exponents reported in \cite{Mario}.
To confirm our results, 
we have measured the excess 
velocity produced by
tilting the initial substrate and
imposing helicoidal boundary conditions \cite{Barabasi}.
The inset in Fig.\ \ref{meanh} shows that this mean velocity is well fitted by
a parabola, as expected for KPZ behavior. It is important to
stress that identical results
are obtained using the jump rules in \cite{Mario}. 

\begin{figure}
\begin{center}
\epsfig{file=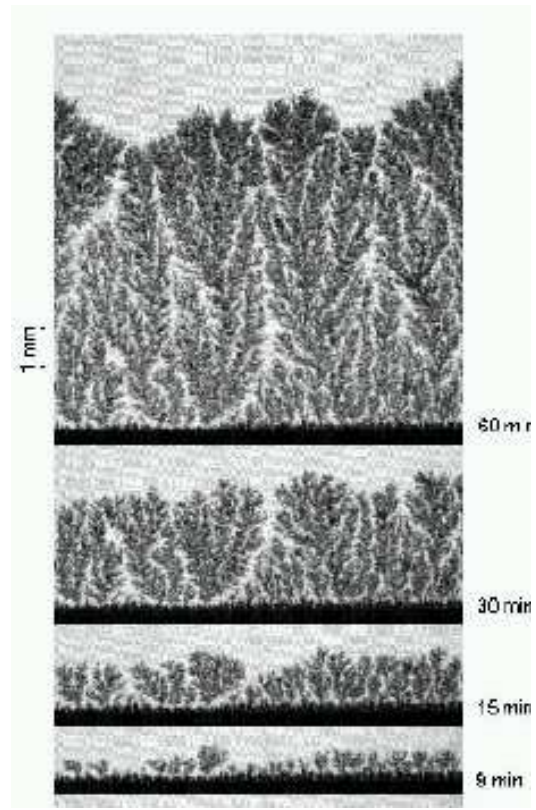, width=4.2in, angle=90}    
\end{center}
\caption{{\em in situ} lateral micrographs showing the interface 
evolution from $t=0$ to $t=60$ min for Ag ECD at $j=1$ mA\,cm$^{-2}$ in
$5\times 10^{-3}$ M Ag$_2$SO$_4 + 10^{-2}$ M H$_2$SO$_4+0.5$ M Na$_2$SO$_4$.
Taken from \protect\cite{Schilardi}
with kind permission from the authors.}
\label{roberto2}
\end{figure}

\begin{figure}
\begin{center}
\epsfig{file=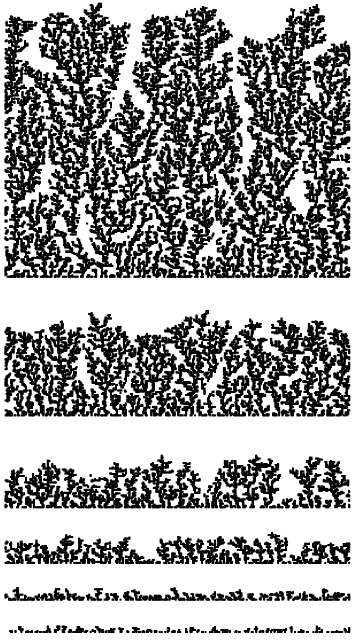, width=1.8in, angle=0}    
\end{center}
\caption{Sequence of snapshots of the evolution of an aggregate 
grown with MBDLA without SD. Parameters are:
$p=0.75$, $s=1$, $r=1$ (i.e., no surface diffusion). Times
(in our units, see text) are (top to bottom):
$35\times 10^6$,
$17.5\times 10^6$,
$8.75\times 10^6$,
$5.25\times 10^6$,
$3.5\times 10^6$, and
$1.75\times 10^6$.}                             
\label{nueva}
\end{figure}

Interestingly, Schilardi {\em et al.} \cite{Schilardi} report 
experiments with large currents 
(equivalent to the large values of the bias $p$) in excellent
agreement with
our model. They observe  
the same three time regimes: An initial 
transient with a behavior which could not be measured
due to the resolution of the
experimental device; a second transient with $\beta>1$ characterized by the
growth of isolated branches, and a third asymptotic regime at which the
interface is characterized by KPZ exponents.  A plot of the
mean interface velocity {\em vs.}\ time is also given, showing a crossover 
from the unstable regime to the stable one in accordance with
MBDLA predictions as can be seen in  Fig.\ \ref{meanh}. 
The global width crosses over from the
instability ($\beta>0.5$) to $W(L,t)\sim t^{1/3}$
at the time pointed out by
the arrow.
Note that MBDLA cannot yield $\beta$ larger than one, because of its discrete
growth rules. This would mean that the interface width 
grows faster than the interface mean height. Finally, the evolution of 
the morphology during the experiment is also the same in MBDLA and
in the experiment, as seen by comparing
Fig.\ \ref{roberto2}, taken from \cite{Schilardi}, and Fig.\ 
\ref{nueva}, obtained in our simulations.

\begin{figure}
\begin{center}
\epsfig{file=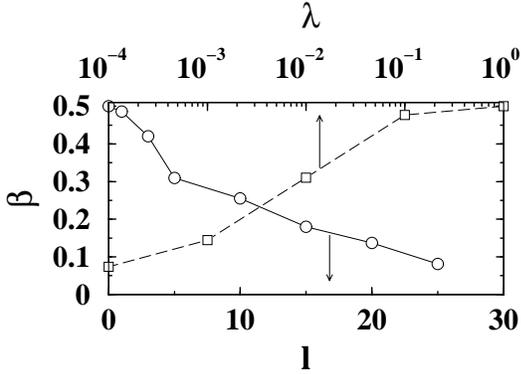, width=2.7in, angle=0}    
\end{center}
\caption{Dependence of the growth exponent $\beta$ on: ($\circ$)
diffusion length $l$ (rule A) 
and ($\Box$) attachment probability $ \lambda$ (rule B).}
\label{scaling_rule_AB}
\end{figure}

We now consider the scaling behavior of MBDLA+SD for the different
diffusion rules.
As we have pointed out in the preceding section, large values of the diffusion
length $l$ (rule A) generate flat aggregates. This
means that $\beta\rightarrow 0$ as $l\to \infty$ at early times. 
Fig.\ \ref{scaling_rule_AB} shows the lack of
universality in the growth 
exponent $\beta$: It can be seen in this plot that $\beta$ decreases
with $l$ as we expected. The same happens with rule B:
As in the case of rule A, the growth exponent $\beta$ depends strongly on the 
attachment probability, $\lambda$ (rule B). As depicted in
Fig.\ \ref{scaling_rule_AB}, the dependence is similar to
that of model A since the diffusion length is proportional to $\lambda^{-1}$.

\begin{figure}
\begin{center}
\epsfig{file=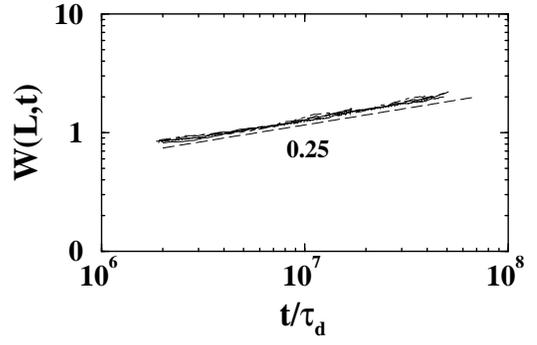, width=2.7in, angle=0}    
\end{center}
\caption{Global width {\em vs} $t/\tau_d$ for $r=0.05$, $0.1$,
$0.15$, $0.2$ and $0.25$,
with $p=0.5$, $s=1$ and $c=0.05$. The dashed line is a guide to
the eye, with
slope $0.25$.}
\label{beta_EW}
\end{figure}

\begin{figure}
\begin{center}
\epsfig{file=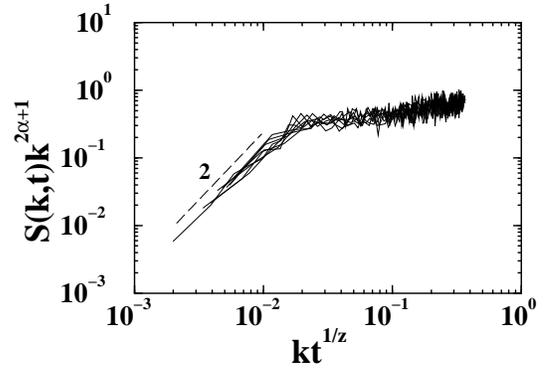, width=2.7in, angle=0}    
\end{center}
\caption{Collapsed power spectrum with $p=2$, $s=1$, $r=0.1$ and
$c=0.05$ using the 
Edwards-Wilkinson universality class exponents at six equally spaced
times from
$8\times 10^6$ to $3\times 10^7$. Dashed line has slope $2\alpha +1=2$.}
\label{EdwardsWilkinson}
\end{figure}

The scaling behavior in MBDLA with SD given by rule C is more complicated.
We can recognize three different kinds of behavior, which we 
summarize as follows: 

$\bf 0.05\leq r\leq 0.25$. 
The characteristic diffusion time is long,
and particles diffuse rather fast along the surface (let us recall 
that they are picked with probability $1-r$ at every Monte Carlo 
trial) without much interaction with particles arriving from the bulk,
thus
yielding compact aggregates,
except if $p\lesssim 0.05$, because then
the Laplacian field creates pillars and grooves. 
After a short transient the global width grows slowly   
and, independently of the applied current, the roughness
exponents are compatible with those of the
Edwards-Wilkinson universality class, whose 
defining equation is \cite{EW}:
\begin{equation}
\label{spde_ew}
\frac{\partial h}{\partial t}=\nu\nabla^2h+ \eta(x,t).
\end{equation}
Figure \ref{beta_EW} shows the global width collapse
obtained 
by rescaling the simulation time.
The plot not only shows  
the Edwards-Wilkinson growth exponent, but also 
the $r$ independence of the results on a wide range of simulation
parameters.
Note that the collapsing time step is the one defined
for MBDLA without SD divided by the characteristic
diffusion time $\tau_d=r^{-1}$.
Figure \ref{EdwardsWilkinson}
shows the collapsed power spectrum
using $\alpha=1/2$ and $z=2$ (and consequently
$\beta=1/4$) consistent with (\ref{Sanom})
with $\theta=0$ for Edwards-Wilkinson
exponents. 
It is important to note that
this kind of dynamic scaling has been observed in
two-dimensional ECD experiments 
\cite{Vazquez}.  
Finally, we have to mention that the restriction 
$r>0.05$ is only due to the extremely long computational 
times needed to study the model for such small values of $r$.

\begin{figure}
\begin{center}
\epsfig{file=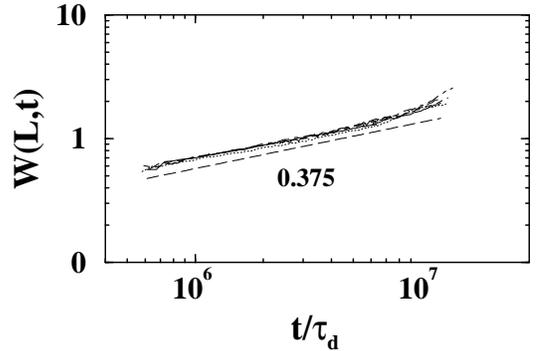, width=2.7in, angle=0}    
\end{center}
\caption{Global width {\em vs.} $t/\tau_d$ for $r=0.35$, $0.40$,
$0.45$, $0.55$ and $0.60$, with
$p=0.5$, $s=1$ and  $c=0.1$. Dashed line is a guide to the eye.}
\label{beta_MBE}
\end{figure}

\begin{figure}
\begin{center}
\epsfig{file=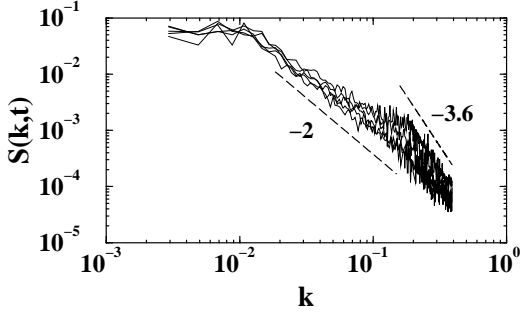, width=2.7in, angle=0}    
\end{center}
\caption{Power spectrum of an interface with $r=0.5$, $p=0.5$, $s=1$
and $c=0.1$ at 
times $10^7, 2\times 10^7, 3\times 10^7, 4\times 10^7,$ and $5\times 10^7$.
Power spectra  are anomalous at short scales. Dashed lines are guides
to the eye.}
\label{espectro_bulto}
\end{figure}

\begin{figure}
\begin{center}
\epsfig{file=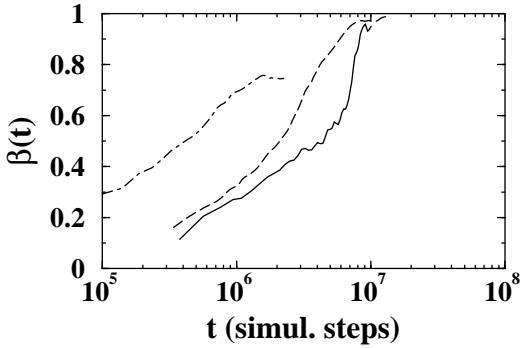, width=2.7in, angle=0}    
\end{center}
\caption{Evolution of growth exponent $\beta$ with time for compact
aggregates with
grooves. Solid line: $p=0.1$ and $r=0.45$; dashed line: $p=0.1$ and
$r=0.5$; and dot-dashed 
line: $p=0.1$ and $r=0.7$.}
\label{Beta_1}
\end{figure}

$\bf0.3\leq r\leq 0.7$. For large $p$, the interface is compact and grows with 
constant velocity. The scaling is similar to that of the preceding case.
When $p\rightarrow 0$, 
initially the interface is rough and the growth exponent $\beta$ is in the 
range $0.35-0.40$ (see Fig.\ \ref{beta_MBE}).  
Some experiments have reported similar interfaces at early stages 
of growth \cite{Tesis_Juanma}: Specifically, they obtained exponents 
consistent with the linear MBE growth model universality class ($\alpha=3/2$, 
$\beta=3/8=0.375$ and $z=4$) that is, their interfaces could be 
described by the equation \cite{Villain}: 
\begin{equation}
\frac{\partial h}{\partial t}=-K\nabla^4h+ \eta(x,t).
\label{mbe}
\end{equation}
Note that for this model $\alpha>1$, so the interfaces generated with
Eq.\ (\ref{mbe}) are super-rough.
In our case this short regime ceases when the mean interface height,
$\bar{h}(t)$, is about $8$ to $10$ monolayers and the global width, $W(L,t)$,
is about $1$. This is compatible with the referred experiments except that 
we do not observe the super-rough power spectrum. 
Actually, in our case the tail of $S(k)$ presents a time shift at 
large wave vectors (Fig.\ \ref{espectro_bulto}) which is incompatible 
\cite{Tail} with the behavior obtained for Eq.\ (\ref{mbe}).  
However, the basic phenomena, such as the value of the effective $\beta$ and 
the onset of the instability, are in good agreement with the experiments.
After this transient, 
the aggregates are still compact and develop some
grooves (see Fig.\ \ref{rule_C}). When these grooves appear, the growth
exponent $\beta$ rises dramatically due to the large slopes
produced between grooves.
Figure \ref{Beta_1} summarizes all this by showing the variation 
of $\beta$ with time. 

\begin{figure}
\begin{center}
\epsfig{file=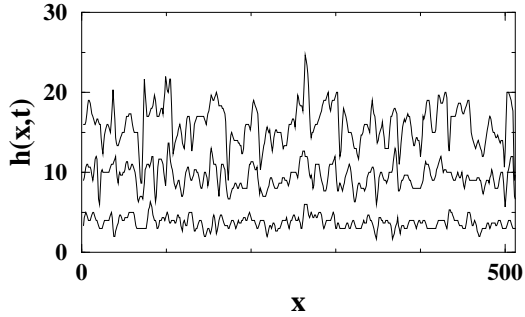, width=2.7in, angle=0}    
\end{center}
\caption{Dynamic evolution of the height $h(x,t)$ with  $p=4$, $r=0.85$,
$s=1$ and $c=0.1$. Snapshots are taken at times $1.2\times 10^7,
2.1\times 10^7,$ and $3\times 10^7$.}
\label{dinamica}
\end{figure}

\begin{figure}
\begin{center}
\epsfig{file=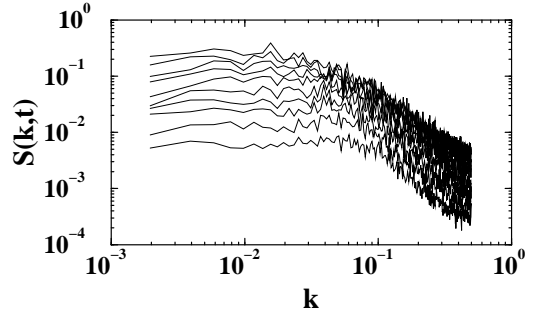, width=2.7in, angle=0}    
\end{center}
\caption{Intrinsic anomalous power spectrum with  $p=4$, $r=0.85$, $s=1$
and $c=0.1$. Lines correspond to interfaces at times $3\times 10^6,
6\times 10^6,$ $9\times 10^6,$ $1.2\times 10^7,$ $1.5\times 10^7,$
$1.8\times 10^7,$ $2.1\times 10^7,$
$2.4\times 10^7,$ $ 2.7\times 10^7,$ and    $3\times 10^7$.}
\label{intrinseco}
\end{figure}

\begin{figure}
\begin{center}
\epsfig{file=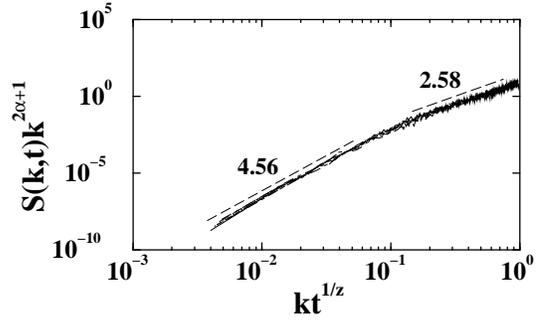, width=2.7in, angle=0}    
\end{center}
\caption{Collapsed power spectrum for the five later curves in  Fig.\
\ref{intrinseco} using: 
 $\alpha=1.78$, $z=2.51$, $\beta=0.71$, and $\alpha_{loc}=0.49$.
Dashed lines show the slope values expected from Eq.\ (\ref{Sanom})
for those exponent values.} 
\label{colapsoS}
\end{figure}

\begin{figure}
\begin{center}
\epsfig{file=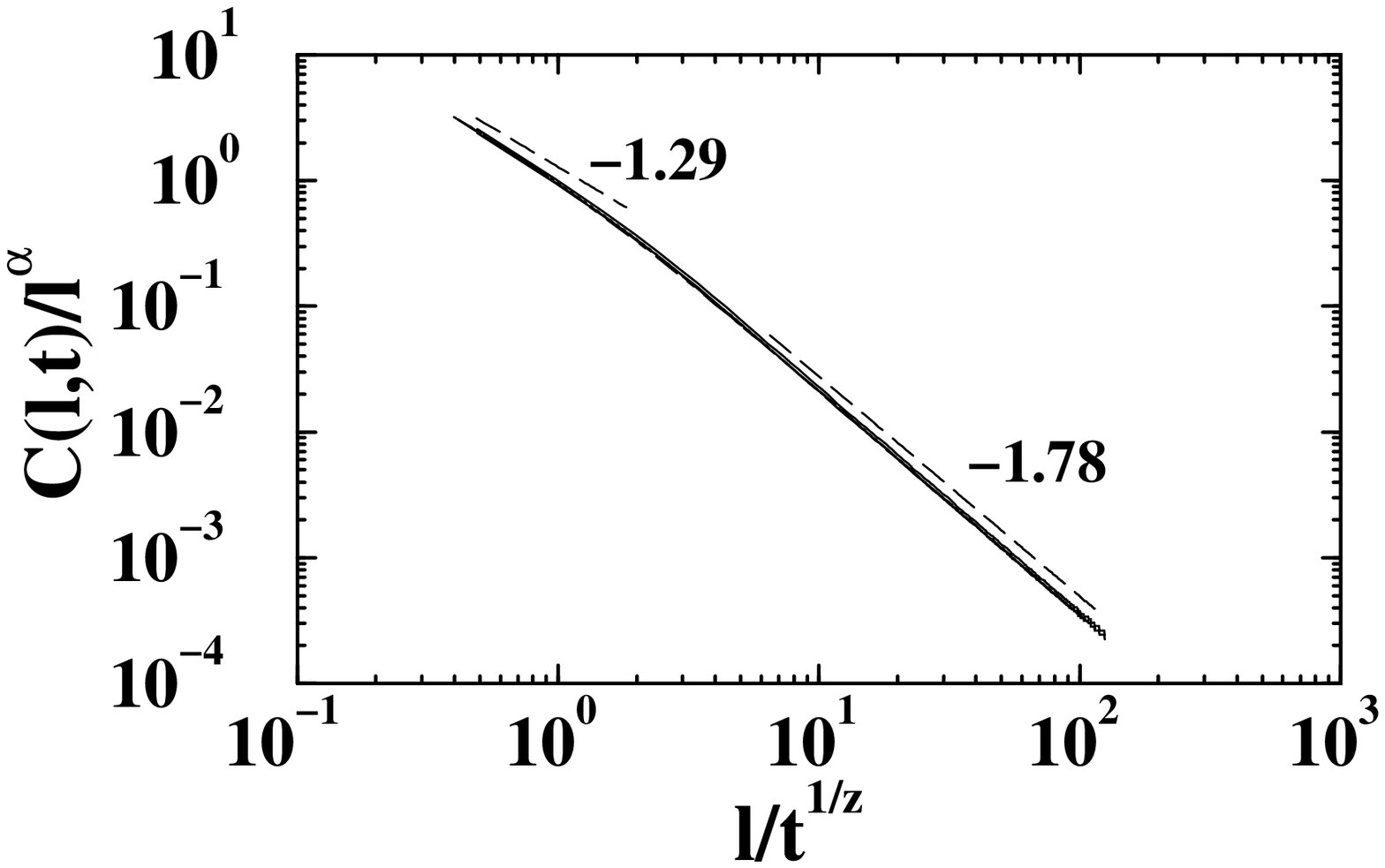, width=2.7in, angle=0}    
\end{center}
\caption{Collapsed correlation function for the same parameters as in
Fig. \ref{colapsoS} for five equally spaced
simulation times using: 
 $\alpha=1.78$, $z=2.51$, $\beta=0.71$, and $\alpha_{loc}=0.49$.
Dashed lines show the slope values expected from Eq.\ (\ref{wlfv})
for those exponent values.} 
\label{colapsoC}
\end{figure}

$\bf0.85\leq r$. Finally, when the diffusion time is short,
three completely different
situations are found 
as a function of the current $p$.
For very large $p$, cations become ballistically driven 
to the aggregate and the unstable transient tends to dissappear (in fact, the 
$p\rightarrow\infty$ limit is the Ballistic Depostion  discrete model, 
which is well known to belong to the KPZ universality class 
\cite{Meakin,Barabasi}). 
When $p\gtrsim 1$ 
the aggregate grows as MBDLA without SD with similar 
parameters, except that in this case the aggregate mean density rises. 
That is, we succesively detect 
a noisy initial transient, the instability associated with the growing
branches, and the KPZ asymptotic limit due to the 
lateral growth of the branches. 
The interfaces within the unstable regime (an example of which
is shown in Fig.\ \ref{dinamica})  
are not self-affine but present intrinsic
anomalous scaling \cite{Lack,Versus}: Figure \ref{intrinseco}
shows the power spectrum for $r=0.85$, $p=4$ and $s=1$. Figure 
\ref{colapsoS} shows the collapse of the  power spectrum and Fig.\
\ref{colapsoC} the collapse of the height-height correlation
function, $C(l,t)$,
achieved in both cases for $\alpha=1.78$, $\alpha_{loc}=0.49$,
$z=2.51$ and $\beta=0.71$.
 
For intermediate $p$ values (between $0.25$ and $1$, for almost every $r$),
the aggregate is formed by several compact thin branches which grow vertically
and parallel to one another.
In this case, the notions of a rough surface or dynamic scaling are 
meaningless.
Finally, for small $p$ some compact branches grow at the expenses of the
others, so typically,
one or two branches grow more than the others. As in
the case of the parallel branches,
it is meaningless to talk about interface roughening.

\section{Discussion and Conclusions}

Our first conclusion is that MBDLA is a simple computational model which 
incorporates in a natural way some of
the basic mechanisms involved in ECD experiments. The original
model \cite{Angel,Mario} was already known to be in good agreement
with some experiments \cite{Bernal}. In this paper, we have provided
much more evidence showing that MBDLA
explains some of the morphological changes due mainly to the applied electric
current and, what is more important, 
it predicts the recently observed KPZ scaling behavior in the
high current limit (for which SD is not too relevant) \cite{Schilardi}
and observed also at low currents \cite{Kahanda}. 
Before this regime is reached, there is an unstable transient within which 
MBDLA interfaces present intrinsic anomalous scaling. We believe this type of
scaling is due to SD not being able to communicate different portions of
the interface fast enough, so that they grow independently from one another.
This is analogous of the anomalous scaling occurring in the non-linear 
surface diffusion equation studied in \cite{Amar}. In our case, the different
portions feature a value of the roughness exponent $\alpha_{loc} \approx 0.5$, 
similarly to the interface subject to columnar disorder studied in
\cite{Lack,Versus}. 

Secondly, the main point of our paper is that, as we have seen,
MBDLA without SD cannot explain low current experiments in which the 
characteristic dense branching aggregates of high current experiments
are replaced by
compact and column-like aggregates. Our working hypothesis was 
that the latter kind of patterns are
due to the competition between the Laplacian field of the
cations in the disolution
and the SD current on the interface. Thus, our ECD model, which we wanted 
to improve as to explain,
at least qualitatively, the complete ECD phenomenology,
should incorporate a new rule
for the diffusion of the adatoms attached to the aggregate.
Hence, we have tried out some SD rules similar to those  
often used in growth models for molecular beam epitaxy \cite{Krug}.
We have verified that instantaneous diffusion rules, namely rules that
``freeze'' the bulk particles while the most recently attached particle
finds its way through the surface, do not lead to correct results in 
the low current limit, and produce very unrealistic, flat-topped 
morphologies. We have then been forced to conclude that 
the nonlocal character of MBDLA demands a diffusion rule which 
couples the overall cation dynamics: This is the rule we have named C.
It introduces a characteristic diffusion time $\tau_d=r^{-1}$ which
competes with the time scale related
to the net flux of particles arriving to the interface
(which, in fact, is proportional to the
applied electric current density). With this SD rule,
the morphologies at low, medium and high currents are
compatible with those observed
by Trigueros {\em et al.} \cite{Sagues} for low, medium and
high applied voltages respectively.
This diffusion time $\tau_d$ cannot be controlled from
the experimental point of
view, but fortunately there are wide ranges of parameters over
which the simulated morphologies hardly change, which means that 
the description of the experiments provided by MBDLA with SD is 
robust and does not need an incontrollable parameter to be tuned.
We have also compared MBDLA with SD with the experiments
reported by Pastor and
Rubio \cite{Tesis_Juanma,Pastor}, which characterize the product
interfaces by the
MBE exponents. MBDLA seems to reproduce the latter behavior for very
short times and short length scales as can be seen in Figs.\
\ref{beta_MBE} and
\ref{espectro_bulto}, but these results are not too significative,
as they are not as accurate as we would need to make any strong
claim, and could be due to
the appearance of a characteristic short length scale. There is a 
difficulty in this respect, as MBDLA scaling is intrinsically anomalous 
whereas the results in \cite{Tesis_Juanma,Pastor} support standard,
super-rough scaling. With the data presently at hand we have to 
conclude that MBDLA with SD does not describe quantitatively all
the aspects of the very low current regime, but the fact it does 
describe most of them and, above all, the compactification of the 
aggregates, makes us confident that MBDLA with SD is a very good 
{\em general} model for ECD. 

To conclude, we note that the model presented has the basic
ingredients of ECD
phenomena: Diffusion, electromigration and surface diffusion,
but for this reason, 
we have to pay a big price in terms of computational
time. MBDLA without SD is a very time-consuming model, and the diffusion
rules make the analysis and the simulations 
a patience exercise. It has certainly been an improvement to find that
SD rule C allows us to skip the sticking probability parameter, thus 
resducing the parameter space, but even then if 
averages of relevant quantities over large ensembles are required, 
a great deal of computational resources would be needed. 
Of course, this disadvantage can be removed by a careful reprogramming
of the algorithm, but that is another line of research. As our goal 
was to identify the most important factors involved in ECD, 
we do believe that 
despite the computing limitations of the model, 
MBDLA with SD is a powerful tool 
to repoduce some unclear features of this kind of growth experiments,
and has helped understand what are the most relevant transport properties
and how they couple in different parameter regions. We hope that this 
work suggests further experiments to find out whether MBDLA with SD is
the complete, general model for ECD or else if there are still regions
which need separate modelling.

\acknowledgments

It is a pleasure to thank
Miguel Angel Rodr\'\i guez and
Francesc Sagu\'es 
for their help through
many discussions on the modeling of ECD and on the subtleties of
dynamic scaling during the last few years, as well as for a critical
reading of the manuscript.    In addition,
we are thankful to Javier Buceta, Juanma Pastor, Javier de la Rubia,
and Miguel Angel Rubio for sharing with us their results and expertise
on ECD at very low currents. We would also like to thank Roberto Salvarezza
and Leticia Schilardi for their kind permission to reproduce
Fig.\ \ref{roberto2} from \cite{Schilardi}.
We are indebted to Jos\'e Cuesta and Ricardo Brito for helpful 
suggestions about the simulation rules and to 
the members of GISC for their interest in this work.  
Work at GISC has been supported by DGES (Spain) Grant No.\ PB96-0119 and
CAM Grant No. 07N/0034/98.

\appendix

\section*{Surface Diffusion Rule B}

In Rule B, surface diffusion starts when a particle has first arrived
to the aggregate and has attached to it (with probability $s$).
The particle jumps with equal probability to one of its two nearest 
neighboring sites on the aggregate until it increases its coordination 
number. The particle has an additional probability, $\lambda$, 
of being permanently attached. This kind of particle is usually termed
a mortal random walker \cite{Hughes}. The random walk is performed 
between two absorbing boundaries, namely, a couple of sites with higher 
coordination (two or three). One could try to determine {\em a priori}
the total number $N$ of jumps the particle has to perform in each realization 
drawing such number from the probability for the particle to take 
$N$ steps on a flat line if it avoids to stick $N-1$ times and 
{\em die} at the $N$th jump. This probability can be easily calculated 
to be given by 
\begin{equation}
P_N(\lambda)=\lambda(1-\lambda)^{N-1}.
\label{lambda_prob}
\end{equation}
However, the absorbing boundaries disallow this procedure.
In any case, we have compared the simulation results by allowing the particle
to perform an actual mortal random walk, and to perform a simple random of 
$N$ steps given by Eq.\ (\ref{lambda_prob}). Both results are hardly different.
Thus, we can approximately calculate from
Eq.\ (\ref{lambda_prob}) the mean and variance of the maximum 
number of jumps, given by:
\begin{eqnarray}
\bar{N}=\frac{1}{\lambda},\\
\sigma_N=\frac{\sqrt{1-\lambda}}{\lambda}.
\end{eqnarray}
For a flat interface, the particle mean position would be $0$ but its variance
would be
\begin{equation}
\sigma_N=\bar{N}^{1/2}/2=1/(2\lambda^{1/2}),
\end{equation}
which provides the characteristic diffusion length $l_D=1/(2\lambda^{1/2})$.

\end{multicols}

\begin{multicols}{2}

\vspace*{2cm} 

\vspace*{1cm}

\end{multicols}


\begin{references}
\bibitem{Comentario_p} We stress that even if the probabilities
	are different
	of those in Refs.\ \cite{Angel} and \cite{Mario}, the drift velocity
	towards the cathode can be shown in all cases to be
proportional to $p$.
\end{references}
\end{document}